# Investigation into Open-Ended Fitness Landscape through Evolutionary Logical Circuits

Masaki SUYAMA[1,2]* and Kosuke SATO[3]*

1 Meiji Gakuin University, Shirokanedai 1-2-37, Minato-ku, Tokyo 108-8636, Japan

2 Keio University, Mita 2-15-45, Minato-ku, Tokyo 108-8345, Japan

3 Meiji University, Kandasurugadai 1-1, Chiyoda-ku, Tokyo 101-8301, Japan

* Author for correspondence (msuyama@ed.meijigakuin.ac.jp or sato_kosuke@meiji.ac.jp).

short title: Investigation of open-ended fitness landscape

(word count: 3010)

**Abstract**

Cumulative cultural evolution is what made humanity to thrive in various ecological and demographic environments. Solutions to the tasks that humans needed to solve could be mapped onto a task space which could take the form of either closed or open-ended fitness landscape, with the former being modeled more extensively than the latter in studies of cultural evolution. In this article, we modified a simulation by Arthur and Polak (2006) that modeled open-ended fitness landscape by using a computer simulation that builds logical circuits with circuits that were built in earlier trials. We used this simulation to clarify the nature of open-ended fitness landscape and to investigate whether the speed of accumulation of culture is increased by an increase in group size. The results indicated that group size increased the speed of accumulation but is limited than expected. Also, when two types of accumulation, invention and improvement, were distinguished the nature of the two differed. In improvement, the trajectory followed a convex function with productivity of one agent decreasing as group size increased. In invention, the trajectory showed a continuous pattern of rapid increase followed by a plateau.

**Introduction**

The accumulation of culture over many generations, or cumulative cultural evolution (Arthur, 2009; Basalla, 1988; Henrich, 2015; Mesoudi, 2011), has led the *Homo* genus to thrive in many, if not all, regions of our planet. Theoretical and empirical investigations into cumulative cultural evolution have paid special interest in cognitive abilities and impact of demography (Dean, Vale, Laland, Flynn, and Kendal, 2013; Henrich, 2004). Although not as frequent, there exist another interest in the fitness landscape (also called adaptive landscape or design space) of cumulative cultural evolution (Acerbi, Tennie, and Mesoudi, 2015; Miton and Charbonneau, 2018). Fitness landscape refers to a set of all potential solutions to a task that agents may produce (Miton and Charbonneau, 2018).

Fitness landscapes could be modeled in two different ways. One that is used above and in most studies of cumulative cultural evolution is *closed* fitness landscape. Closed fitness landscape refers to a fitness landscape that cannot be altered by the decisions or the actions of an agent solving a task. This implies that all the choices in the fitness landscape can be chosen at any time period. For example, Mesoudi and O'Brien (2008) created a fitness landscape that consisted of 4 dimensions (width, height, thickness, and shape), which was used to create a virtual arrowhead. Each dimension was designed with a particular fitness function so that each coordinate in the dimension corresponded to a particular fitness. Participants in their experiment changed the coordinates in each dimension to achieve the largest fitness or calories. Closed fitness landscapes are useful because they are easy to formulize. On the other hand, it also gives participants the opportunity to find the maximum fitness in the early stages of transmission by mere chance.

In contrast, *open-ended* fitness landscape refers to a situation where an accumulation opens up a new landscape for selection to operate. As a result, these cultural evolutionary systems tend to increase in complexity not only because the fitness landscape tend to increase in space, but also these newly opened fitness landscapes can replace the highest peak in the fitness distribution (Clark 1985; Arthur 2009; Solé, Ferrer-Cancho, Montoya, and Valverde, 2002). Usually, the complexity is increased by recombining already created artifacts together to create new ones, which captures the essence of technological evolution (See SI:1 for examples). Derex and Boyd (2016) for instance found experimentally that in an open-ended fitness landscape, increase in group size alone will not increase the chance of finding the optimal solution which built upon the idea formally believed in cultural evolution studies.

Another experiment aimed at modelling the nature of technological evolution by simulating the evolution of logical circuits. Arthur and Polak (2006) simulated a situation where an agent randomly wired together logical circuits to create new circuits. They differentiated newly made circuits by improvement and invention where the former implying circuits that improved preexisting circuits that were made previously and the latter implying circuits that served a new purpose. Logical circuits have a characteristic where key circuits such as AND circuit or OR circuit be used to create diverse number of other functional circuits (see SI:2 for more details). Whenever such inventions were made, these opened newly fitness landscapes that either made previously created circuits more efficient (termed gales of destruction; Schumpeter, 1911) or made newly inventions be invented rapidly (termed technological Cambrian explosion).

This study was able to model open-ended fitness landscape by actually modelling technology that is often invented in industries. Considering the technological evolution often modeled in cultural evolutionary studies lack external validity (Miton and Charrboneau, 2019), models made by Arthur and Polak (2006) could be used to test the theories that were already created in cultural evolution studies in closed fitness landscape.

In the present study, we attempt to further clarify the feature of open-ended fitness landscape by varying group size in the model created by Arthur and Polak (2006). We added conditions where agents in the same society were able to use circuits built by other agents. Just as in Derex and Boyd (2016), we were interested in how various group sizes perform in an open-ended fitness landscape created by Arthur and Polak (2006). Since the present study have no transmission error, we are not particularly concerned about whether increase in group size stop the deterioration of cumulative culture. Additionally, by using Arthur and Polak (2006) simulation, we may be able to see whether group size have different effect on invention and improvement. In the following simulation, agents in the same trial did not interact with one another that could create a synergetic interaction for simplicity.

**Method**

The simulation was a modified version of Arthur and Polak (2006). In the original simulation, several NAND circuits were randomly wired together in a non-cyclic way to make a new circuit that could be used to create another circuit in further trials. This sequence was repeated several thousand times, which created circuits that were often used in reality (e.g. OR circuit, AND circuit, and n-bit ADDER). The simulation used in this experiment added agents that created circuits simultaneously to vary group size.

In the first trial of all conditions, agent(s) started only with a NAND circuit. Agents wired several NAND circuits to create a new circuit that served a new functionality. The minimum and the maximum number of total NAND circuits that could be wired together were 2 and 12 respectively throughout the trials. The new circuit that was created in the first trial was automatically stored in the pool, which was a group of circuits that could be used as a component for making a new circuit in further trials. Each circuit was insured to be a directed acyclic graph. The choice of using which preexisting circuit was determined by a choice function (Arthur and Polak, 2006; SI:2) that specifies probabilities of selection.

Circuits were evaluated by its functionality, i.e. truth table. Preceding the simulations, goals were defined which consisted of specific input-output circuitry (Table 1; SI:3). When the created circuit either met the goal for the first time or was close to meeting the goal determined by the prespecified truth table, the created circuit was called *invention*.

[INSERT TABLE 1 HERE]

Determined by the truth table, when the created circuit met the same functionality as with the circuit that was included in the pool but with less cost (here cost refers to the total number of NAND circuits used since all the created circuits are created from NAND circuits), the created circuit was called *improvement*. When improvement was made, the older circuit that fulfilled the same functionality was deleted from the pool. On the other hand, if the circuit was neither an invention or an improvement, the circuit was called *junk* and was never included in the pool.

The conditions were separated by how many agents were involved in creating the circuits in the simultaneous trial. The group sizes were 1, 2, 4, and 8. In the conditions that had multiple agents, agents created circuits by themselves. After creating the circuits, the circuits were pooled together. The pooled circuits could be used by all agents in the next trial. This meant that the agents had no synergetic influence on one another.

In each condition, 1 replication consisted of 100,000 trials and 20 replications were run. Whenever all the goals were met, the replication was terminated. The program was created in GNU CLISP (ver. 2.49) which is an implementation of Common Lisp. The simulation was run on 16GB memory Windows 10.

**Results**

The number of trials by conditions and replications are shown in Table 2. Replications were terminated when all the given goals had been achieved. Replication 16 of group size 4 was aborted by memory error, thus we excluded it from the following results. Since the main results are on inventions and improvement, results for goals and junks are in SI:4.

[INSERT TABLE 2 HERE]

**Basic properties of evolution in group size-1**

Since the results of size-1 condition were identical to Arthur and Polak (2006), we regarded the results of size-1 condition as a baseline. The primary results of group size-1 are shown in Figure 1.

[INSERT FIGURE 1 HERE]

The number of inventions in each trial is shown in figure 1a. Generally, we saw a repeated pattern where there was a rapid increase in invention followed by a period where there was minimal increase, which resembled that of a continuous convex function. This continuous sigmoidal like pattern was present in all replication.

Figure 1b shows the number of improvements that made the circuit more efficient. In total, we saw a general increase in improvements as trials progress. Though in some replications we saw a repeated convex like shape in the increase in improvement, compared to invention however, the pattern was not robust.

**Comparison between group sizes**

Figure 2 shows the comparison within invention. The average number of inventions indicated that as group size increased, cumulative increase in invention started to resemble that of a repetitive sigmoidal shape. This indicated that as group size increased, the gentle slope seen in replications under group size-1 was quickly followed by a rapid increase. And as quicker the goals were met, the plateau in bigger group sizes became longer along the end of the simulation. To roughly compare the difference between group sizes, we compared with the baseline by speeding up group size 1 times the number of group size compared (Figure 2b). Similar to goals, the light line was above the solid line indicating that the productivity of group size was lower than expected. We also fitted the data using OLS with the power function applied through the curve_fit function from SciPy (ver. 1.2.1) module in Python (ver. 3.6.8). Since the data fitted poorly in invention, it suggested that the increase in invention did not follow a power function and those follow a more sophisticated one.

**[INSERT FIGURE 2 HERE]**

Figure 3a shows the actual data and their average results from improvement in each condition. Consistent with inventions, speed of improvements also increased with increase in group size similar to a gentle upward convex function. When the speed of group size-1 was increased to be compared with other group sizes, the two lines seemed to overlap with one another. This indicated that the speed of improvements was proportional to that of group size. Just as in invention, we fitted the data with OLS. The results indicated that improvement was roughly proportional to the square root times the group size, but there was a pattern in which the estimated value of the index increased as the group size increased. This suggested that as the group size increased, the rate of increase of the slope became steeper. The results from OLS also suggested that the properties of invention and improvement differed. We compared the normalized RMSE (Root Mean Squared Error) from the OLS between invention and improvement in Figure 4. This suggested that improvement fit well with the OLS more than invention indicating that the nature of the two differed.

**[INSERT FIGURE 3 HERE]**

**[INSERT FIGURE 4 HERE]**

**Discussion**

Open-ended fitness landscape is one of the features of technological evolution. However, the nature of this aspect is understudied especially in fields like cultural evolution where many have pointed out the mechanisms that facilitate the growth in technology under a closed fitness landscape. The main goal of this study was to examine the architecture of open-ended fitness landscape through the evolution of logical circuits with varying group sizes, which is one of the mechanisms identified as the cause of cumulative cultural evolution (Henrich, 2004; Henrich, 2017).

Results showed that the increase in speed of invention by increase in group size was lower than the baseline (which was the n-times the speed of group size-1) and the way that inventions accumulated were similar to a repetitive sigmoidal function. On the other hand, improvement matched that of the baseline and the rate in which improvement increased was square root times the group size, which means that the effect of group size becomes smaller as group size becomes bigger.

One reason for the decrease in the effect of group size in improvement could be due to a chance that one of the agents in the group created an improvement so efficient that other agents could not improve any further. The chance that any agent creates an efficient circuit increase as group size increase, which is similar to the point seen in Henrich (2004).

The interesting finding is that the function of invention and improvement differed. Kaufmann (1993) has argued that when agents start to hill climb in closed fitness landscape, the probability that the agent can climb further decreases as she reaches the top. This would be an explanation for marginally decreasing function in improvement. In invention, we observed an s-shaped curve, which could be seen in many field research (Foster 1986; Christensen 1997; Christensen 2009). As mentioned by Arthur and Polak (2006), the rapid take-off of invention is subject to goals (e.g. AND circuit, OR circuit, etc.) being met. This means that when a goal is met, inventions using that goal circuit rapidly increase. However, at some point, the limits of using that goal circuit are reached and the growth of inventions stops. Such a result was never reported in cultural evolutionary models and can be considered as a key finding in this study. Nonetheless, since this is a post hoc analysis, whether or not the results that inventions take the form of a repetitive s-curve is open for debate.

Besides the difference between invention and improvement, one of the take-home messages is that even if group size increases, the productivity of one agent being added decreases as group size becomes bigger. This also suggests that an increase in group size is sufficient to maintain technology but not enough to accelerate the speed of innovations. However, we did not include synergetic interactions for simplicity in this study which means there is still room to argue that with interaction, group size does increase the speed of innovations. On the other hand, behavioral sciences have shown that group processes do not always have a positive effect (e.g. groupthink, social loafing). We need further examination to see whether interaction does increase the speed of innovations in an open-ended fitness landscape.

There are still many candidates that may affect the speed of innovations. One example is network structure (Frenken 2006; Powell et al. 1996). Derex and Boyd (2016) have reported experimentally that partial connectivity increases the innovations of a group more than a full connected group in an open-ended fitness landscape. Such mechanisms are needed to be explored in future research.

## Acknowledgements

We would like to thank the editor and the reviewers for taking the time to review our work. We would also like to thank Dr. Jackson Polak for sharing with us the program used in Arthur and Polak (2006). This study was supported by JSPS KAKENHI grant-in-aid for JSPS fellows #15J02148 to Masaki Suyama.

## Author Contributions

The initial research was designed and conducted by M.S. and K.S. K.S. programmed and analyzed the data. M.S and K.S. wrote the manuscript.

**Table 1 Goals that were defined preceding the simulations.**

Caption: There were 16 goals in total. N-bit adder had a range of 1 to 8.

| Name | Inputs | Outputs |
|---|---|---|
| NOT | 1 | 1 |
| IMPLY | 2 | 1 |
| AND | 2 | 1 |
| OR | 2 | 1 |
| XOR | 2 | 1 |
| EQUIV | 2 | 1 |
| 3-WAY-AND | 3 | 1 |
| FULL-ADDER | 3 | 2 |
| $n$-BIT-ADDER | $2n$ | $n+1$ |

**Table 2 The number of trials by conditions and replications.**

Caption: Replications were terminated automatically when all prespecified goals were met. Dashed data represents the data that was terminated due to error.

|             | Conditions |         |         |         |
| ---         | ---        | ---     | ---     | ---     |
| Replication | size-1     | size-2  | size-4  | size-8  |
| 1           | 100,000    | 100,000 | 95,950  | 42,000  |
| 2           | 100,000    | 100,000 | 58,000  | 75,216  |
| 3           | 100,000    | 100,000 | 100,000 | 53,636  |
| 4           | 100,000    | 100,000 | 88,867  | 51,952  |
| 5           | 100,000    | 100,000 | 100,000 | 38,965  |
| 6           | 100,000    | 100,000 | 100,000 | 57,881  |
| 7           | 100,000    | 97,981  | 100,000 | 37,000  |
| 8           | 100,000    | 100,000 | 90,000  | 62,796  |
| 9           | 100,000    | 100,000 | 100,000 | 66,986  |
| 10          | 100,000    | 100,000 | 68,988  | 38,100  |
| 11          | 100,000    | 90,974  | 93,754  | 44,977  |
| 12          | 100,000    | 100,000 | 79,970  | 66,716  |
| 13          | 100,000    | 100,000 | 100,000 | 44,000  |
| 14          | 100,000    | 100,000 | 100,000 | 56,974  |
| 15          | 100,000    | 100,000 | 75,949  | 59,000  |
| 16          | 100,000    | 100,000 | ~~25,965~~ | 49,989 |
| 17          | 100,000    | 100,000 | 79,399  | 38,717  |
| 18          | 100,000    | 100,000 | 83,987  | 77,000  |
| 19          | 100,000    | 100,000 | 64,000  | 44,617  |
| 20          | 100,000    | 100,000 | 100,000 | 46,772  |

**Figure 1 The cumulative score for each variable in group size-1.**

Caption: Cumulative score for each replication in group size-1. (a) represents the results from invention and (b) represents the results from improvement.

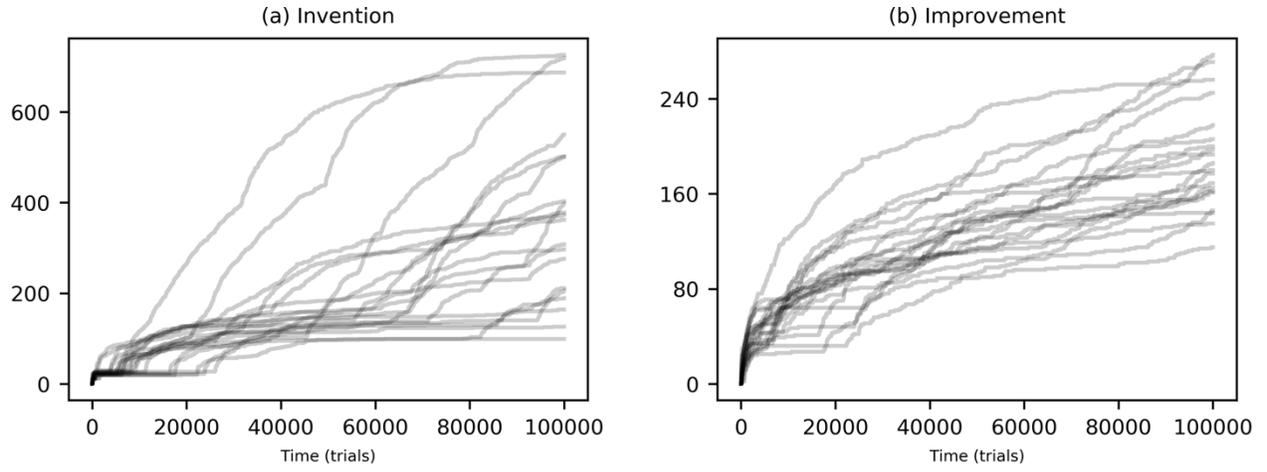

**Figure 2 The cumulative score for invention in each condition.**

Caption: (a) Light lines indicate raw data from each replication. Solid lines indicate the average. The average value was displayed up to a point where no termination was present in all trials. (b) Solid lines represent the average cumulative score. Light lines represent the prediction calculated with group-size 1. (c) Solid lines represent the average cumulative score. Light lines represent the fitted line using power-law function.

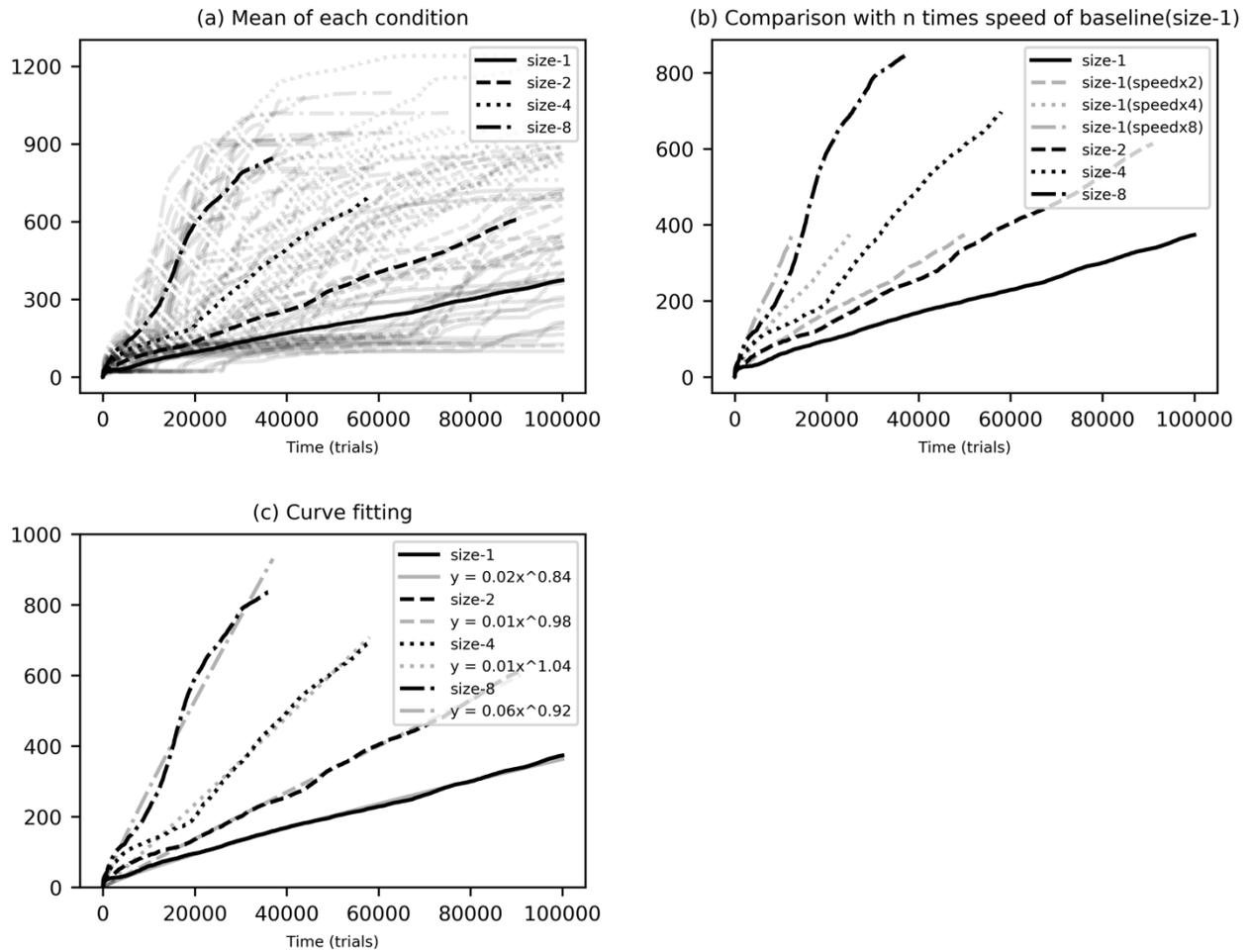

**Figure 3 The cumulative score for improvement in each condition.**

Caption: (a) Light lines indicate raw data from each replication. Solid lines indicate the average. The average value was displayed up to a point where no termination was present in all trials. (b) Solid line represents the average cumulative score. Light lines represent the prediction calculated with group-size 1. (c) Solid lines represent the average cumulative score. Light lines represent the fitted line using power-law function.

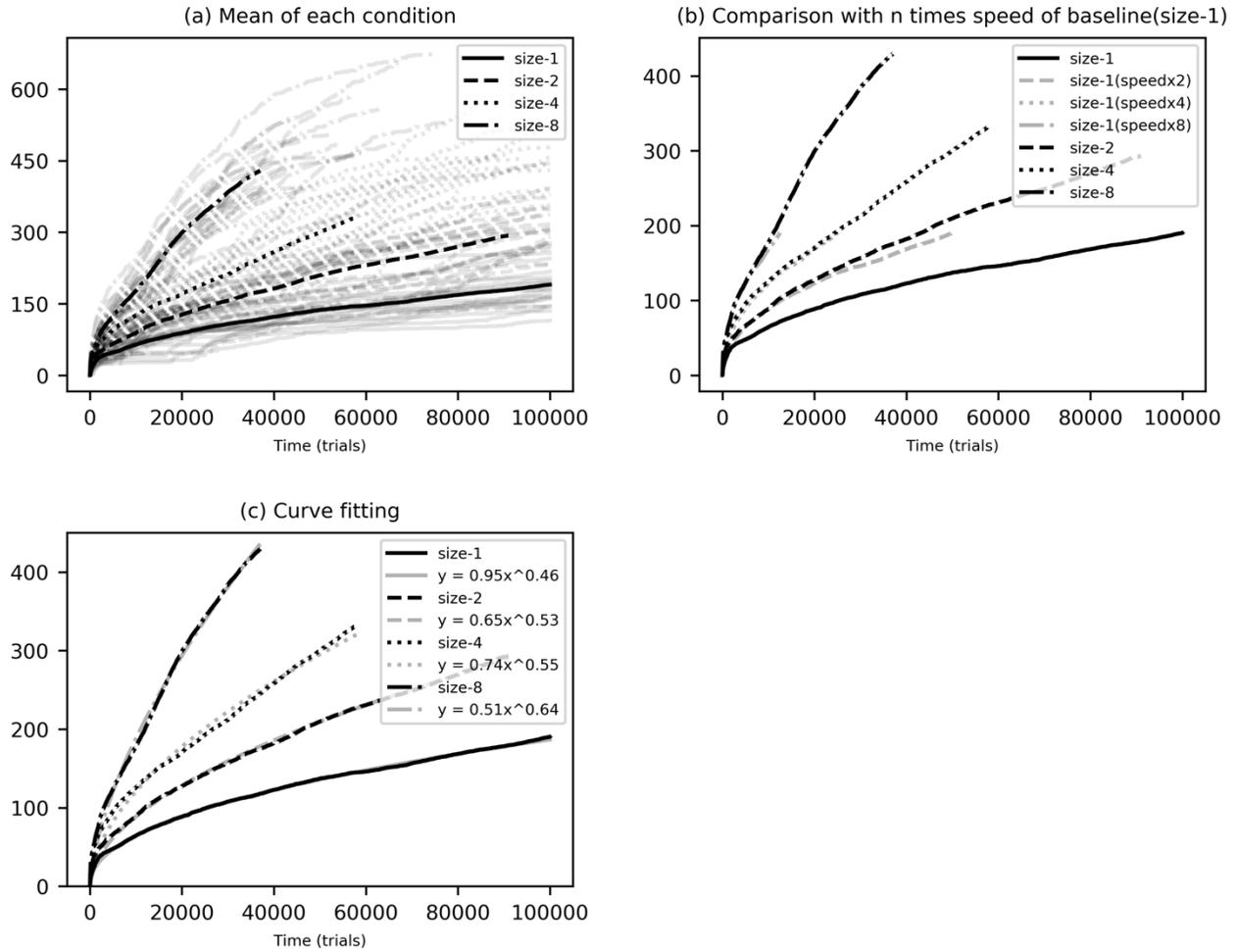

**Figure 4 Comparison of normalized RMSE from the fitted model in all condition.**

Caption: Bar graph representing the normalized RMSE (Root Mean Squared Error) from the curve fit model between invention and improvement.

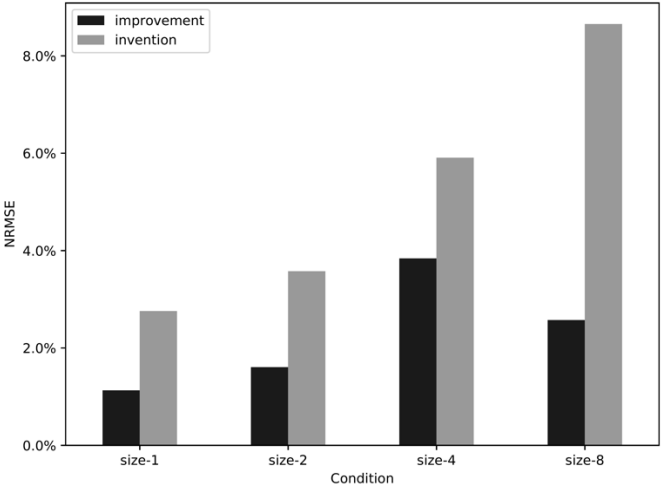

**SI: 1 Examples of open-ended fitness landscape**

For example, Solé, Valverde, Casals, Kauffman, Farmer, Eldredge (2013) have researched the relationship between patents. Since patents need to report other patents that they were influenced by, network structure of the relationships between patents can be easily defined. The network structure showed a combinatorial nature of patents whereby older and similar patents were recombined to make very different patents. Such emphasis on open-ended fitness landscape can be found in fields such as management (Henderson and Clark 1990; Frenken 2006), complexity theories (Solé et al. 2002), and even in language evolution (Zipf 1949; Corominas-Murtra et al. 2018) but not as much in cultural evolution studies of technology (exceptions can be found in Derex and Boyd, 2016). This might be the case because open-ended fitness landscape is hard to model considering its combinatorial nature.

## SI: 2 Choice function

In the simulation, two circuits were randomly wired together by a choice function. In the choice function, one of the options listed below were selected determined by given probability.
1) Constant (true or false)
2) Initial value (NAND circuit)
3) Select a circuit from the pool by equal probability. However, ones that are replaced by improvement will not be included.

## SI: 3 Explanations of key circuits and their structure using NAND circuit

Some of the key circuits or goals are listed here with its truth table and its circuitry using NAND circuits.

## SI Figure 1-1 NOT circuit

|     | Input | Output |
|-----|-------|--------|
| NOT | 1     | 0      |
|     | 0     | 1      |

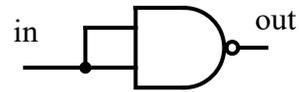

## SI Figure 1-2 IMPLY circuit

|       | Inputs | | Output |
|-------|---|---|---|
|       | A | B | Y |
| IMPLY | 0 | 0 | 1 |
|       | 1 | 0 | 1 |
|       | 0 | 1 | 0 |
|       | 1 | 1 | 1 |

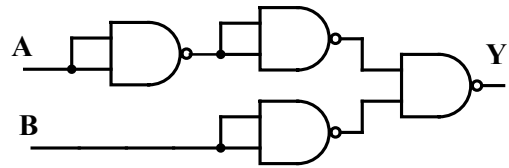

## SI Figure 1-3 AND circuit

|     | Inputs | | Output |
|-----|---|---|---|
|     | A | B | Y |
| AND | 0 | 0 | 0 |
|     | 1 | 0 | 0 |
|     | 0 | 1 | 0 |
|     | 1 | 1 | 1 |

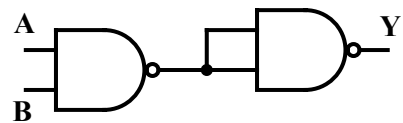

## SI Figure 1-4 NOR circuit

|     | Inputs | | Output |
|-----|---|---|---|
|     | A | B | Y |
| XOR | 0 | 0 | 0 |
|     | 1 | 0 | 1 |
|     | 0 | 1 | 1 |
|     | 1 | 1 | 0 |

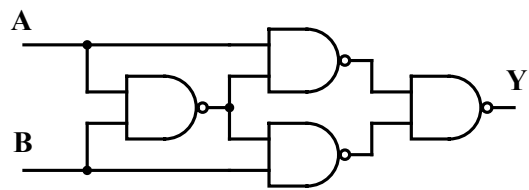

## SI Figure 1-5 EQUIV circuit

|       | Inputs | | Output |
|-------|--------|---|--------|
|       | **A**  | **B** | **Y** |
| EQUIV | 0      | 0 | 1      |
|       | 1      | 0 | 0      |
|       | 0      | 1 | 0      |
|       | 1      | 1 | 1      |

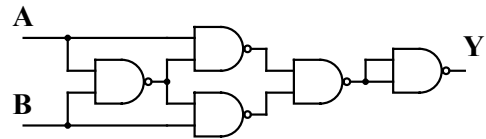

## SI Figure 1-6 OR circuit

|    | Inputs | | Output |
|----|--------|---|--------|
|    | **A**  | **B** | **Y** |
| OR | 0      | 0 | 0      |
|    | 1      | 0 | 1      |
|    | 0      | 1 | 1      |
|    | 1      | 1 | 1      |

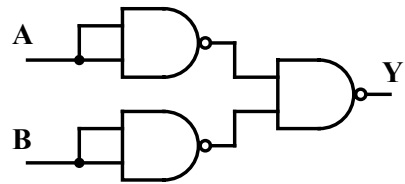

## SI Figure 1-7 FULL-ADDER circuit

|             | Inputs | | | Outputs | |
|-------------|---|---|------|-------|---|
|             | **A** | **B** | **C-in** | **C-out** | **S** |
| FULL-ADDER  | 0 | 0 | 0 | 0 | 0 |
|             | 0 | 0 | 1 | 0 | 1 |
|             | 0 | 1 | 0 | 0 | 1 |
|             | 0 | 1 | 1 | 1 | 0 |
|             | 1 | 0 | 0 | 0 | 1 |
|             | 1 | 0 | 1 | 1 | 0 |
|             | 1 | 1 | 0 | 1 | 0 |
|             | 1 | 1 | 1 | 1 | 1 |

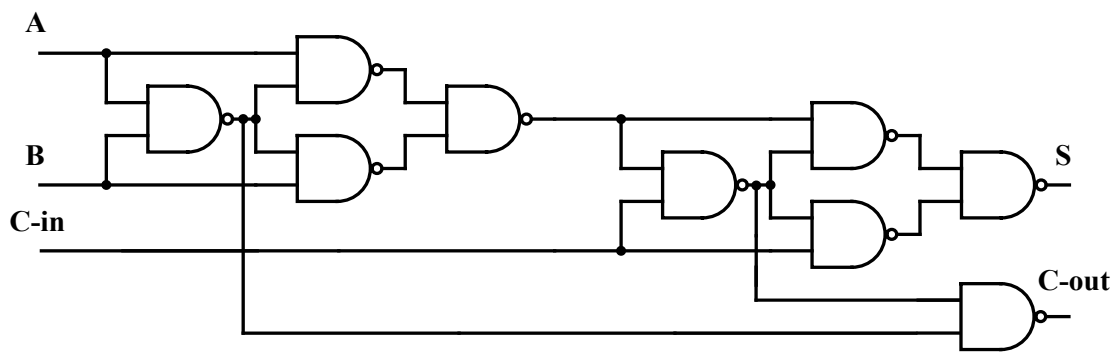

## SI:4-1 Basic properties of evolution in group size-1 (goals and junk)

SI Figure 1a shows the timing of when goals were achieved in each trial. While the trial with the fastest progress reached the 16th goal in about 30,000 trials, the trial with the slowest progress did not yet reach the 11th goal even after 100000 trials. This indicated that a large variation existed when only 1 agent was modifying the circuits.

SI Figure 1b shows the number of cases where neither invention nor improvement occurred. The figure indicated that there was little variance within junk. This was because the number of times innovation and improvement occurred was less than the number of trials.

**SI Figure 1 The cumulative score for each variable in group size-1.**

Caption: Cumulative score for each replication in group size-1. (a) represents the results from goals and (b) represents the results from junk.

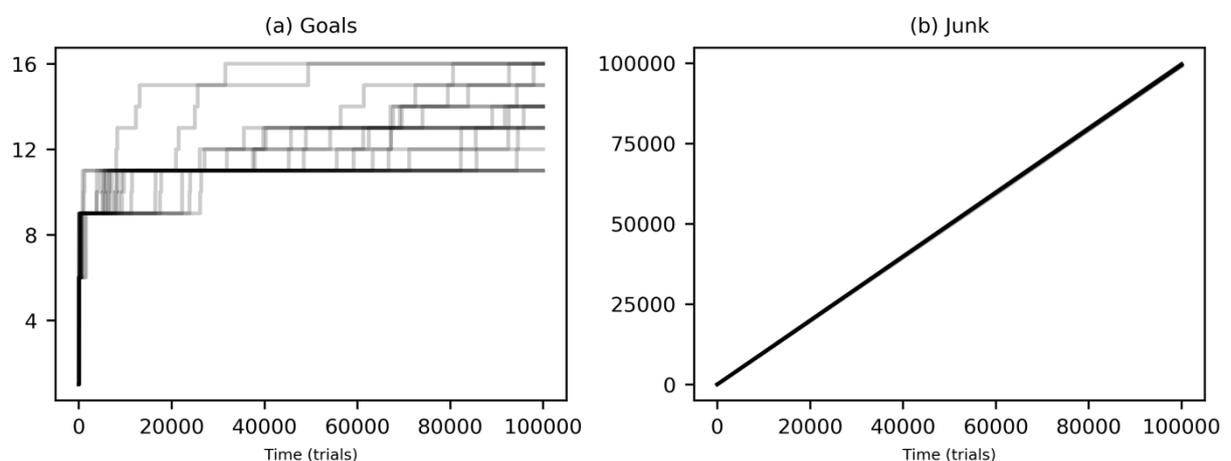

## SI: 4-2 Results for goals

SI Figure 2a shows the number of goals achieved in each condition with the light line indicating the actual data and the solid line the average. The average value is displayed up to a point where no termination was present in all trials. The pattern indicated that as group size increased, so did the speed of goals achieved. Likewise, the difference between conditions became larger as trials advanced. In SI figure 2b, light lines represent data where the speed of size-1 was increased by the number of group size compared (which also could be interpreted as dividing the number of trials in group size-1 by the number of group size compared). Since the light line was above the solid line (which is the average of the actual data) in all conditions, this suggested that productivity of group size was lower than expected.

**SI Figure 2 The cumulative score for goals in each condition.**

Caption: (a) Light lines indicate raw data from each replication. Solid lines indicate the average. The average value was displayed up to a point where no termination was present in all trials. (b) Solid lines represent the average cumulative score. Light lines represent the prediction calculated with group-size 1.

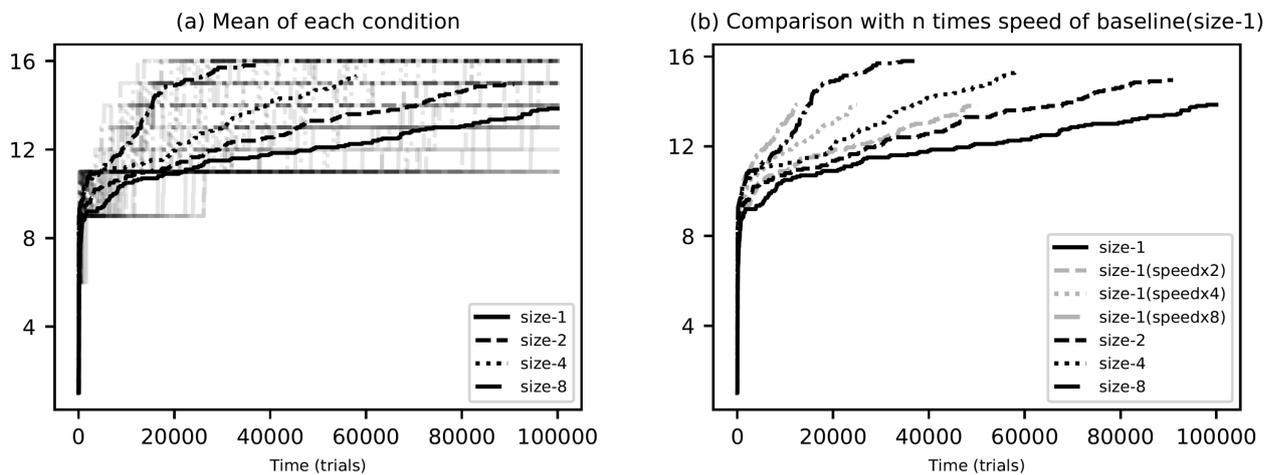

# SI: 4-3 Result for junks

SI Figure 3 shows the results from junk. The accumulation of junk increased as the group size increased (figure 3a). When the speed of group size-1 was increased, the results overlapped with the increased data set. This means that productivity of junk increases proportionally to group size.

**SI Figure 3 The cumulative score for junk in each condition.**
Caption: (a) Light lines indicate raw data from each replication. Solid lines indicate the average. The average value was displayed up to a point where no termination was present in all trials. (b) Solid lines represent the average cumulative score. Light lines represent the prediction calculated with group-size 1.

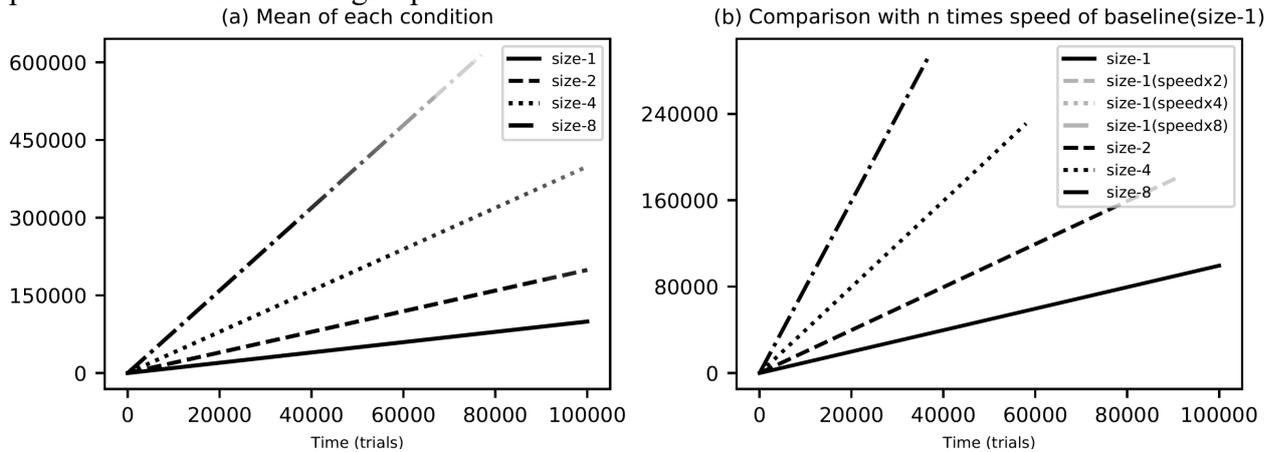